\documentclass[]{aa} 
\usepackage{amsmath,amssymb}
\usepackage{amstext}
\usepackage{natbib}
\usepackage{txfonts}
\usepackage{graphicx}
\usepackage{arydshln}
\usepackage {xcolor}
\usepackage{url}
\bibpunct{(}{)}{;}{a}{}{,} 

\def\m2s2{\hbox{\,m$^{2}$\,s$^{-2}$}} 

\def\chisq{\mbox{$\chi^2$}}

\def\target{WD~1856+534}
 
\usepackage{color}

\begin{document}

\title{A transmission spectrum of the planet candidate WD~1856+534~b and a lower limit to its mass\thanks{Based on observations made with the Gran Telescopio Canarias (GTC), on the island of La Palma at the Spanish Observatorio del Roque de los Muchachos of the IAC, under Director's Discretionary Time GTC2020-144.}}

\author{Alonso, R. \inst{1,2}
\and Rodr\'\i guez-Gil, P. \inst{1,2}
\and Izquierdo, P. \inst{1,2}
\and Deeg, H.~J. \inst{1,2}
\and Lodieu, N. \inst{1,2}
\and Cabrera-Lavers, A. \inst{1,3}
\and Hollands, M.~A. \inst{4}
\and P\'erez-Toledo, F.~M. \inst{3}
\and Castro-Rodr\'\i guez, N. \inst{3}
\and Reverte Pay\'a, D. \inst{3}
}


\institute{
Instituto de Astrof\'\i sica de Canarias, E-38205 La Laguna, Tenerife, Spain
\and Departamento de Astrof\'isica, Universidad de La Laguna, 38206 La Laguna, Tenerife, Spain\label{La Laguna}
\and Gran Telescopio Canarias (GTC), E-38712, Bre\~na Baja, La Palma, Spain
\and Department of Physics, The University of Warwick, Coventry, CV4 7AL, UK
}

\date{Received 15 January 2021 / Accepted 29 March 2021}
\abstract
{The cool white dwarf \target\ was found to be transited by a Jupiter-sized object with a mass at or below 14 M$_{\rm{Jup}}$. We used the GTC telescope to obtain and analyse photometry and low resolution spectroscopy of six transits of \target\,b, with the intention to derive the slope of the transmission spectrum, towards an eventual detection of Rayleigh scattering of the particles in its atmosphere. Such a slope, assuming a cloud-free atmosphere dominated by Rayleigh scattering, could be translated into an estimation of the mass of \target\,b. However, the resultant transmission spectrum is essentially flat, and therefore permits only the determination of lower mass limits of 2.4~M$_{\rm{Jup}}$ at the 2-$\sigma$ level, or 1.6~M$_{\rm{Jup}}$ at 3-$\sigma$. These limits have implications for some of the proposed formation scenarios for the object. We elaborate on the potential effects of clouds and hazes in our estimations, based on previous studies of Jupiter and Titan. In addition, we detected an H$\alpha$ absorption feature in the combined spectrum of the host white dwarf, that leads to the assignation of a DA classification and allows derivation of an independent set of atmospheric parameters. Furthermore, the epochs of five transits were measured with sub-second precision, which demonstrates that additional objects more massive than $\approx$5~M$_{\rm{Jup}}$ and with periods longer than $O(100)$ days could be detected through the light travel time effect.}


\titlerunning{A transmission spectrum of \target\ b}
\authorrunning{Alonso et al.}

\maketitle

\section{Introduction}
\label{sec:intro}
White dwarfs (WDs hereafter) are the final product of the evolution of stars with masses between one and eight solar masses. In the last decades, growing evidence has been obtained about the presence of planetary material around some of these objects in the form of debris disks (e.g. \citealt{Farihi:2016aa, Rebassa-Mansergas:2019aa}) and/or metal pollution in their atmospheres (\citealt{Zuckerman:2003aa,Zuckerman:2010aa,Koester:2014ab}). 

With the Kepler space telescope \citep{Borucki:2010aa},  \citet{Vanderburg:2015ab} found disintegrating material transiting the hot WD~1145+07. These transits exhibit varying depths, shapes, and periodicities in the 4.5-4.9~h range (e.g.~\citealt{Alonso:2016aa,Gary:2017aa,Izquierdo:2018aa}), which is compatible with transiting clouds due to the evaporation of drifting asteroid fragments orbiting the WD \citep{Rappaport:2016aa}. \citet{Vanderbosch:2020aa} found another similar case, but with a much longer period of $\approx$107.2~d, in ZTF~J013906.17+524536.89.

\begin{table*}
\centering
\caption{Log of the GTC observations}
\begin{tabular}{llcccc}
\hline
\hline
Date  & Instrument & Configuration & \#exp. [exp. time (s)] & Seeing (\arcsec) & Res. scatter (\%)  \\
\hline
2019-10-21\tablefootmark{a} & OSIRIS &Sloan $g$, frame transfer & 419[10] & 1.0-3.0 & 0.6 \\
2020-03-02 & OSIRIS &Sloan $i$, frame transfer & 600[5]& 1.5 & 0.4 \\
2020-05-03\tablefootmark{b} & OSIRIS &   R1000R, 0.8\arcsec and 2.5\arcsec slits& 2[530]+13[84]& 1.6-4.0  & 2.5\\
2020-05-10\tablefootmark{c} & OSIRIS &  R1000B, 2.5\arcsec slit& 31[75] & 0.8 & 0.3  \\
2020-05-20 & OSIRIS & R1000B, 2.5\arcsec slit& 43[75] & 0.9 & 0.15 \\
2020-06-13 & EMIR & $J$ filter, STARE, sky bef. and aft. & 45[10]+120[10]+60[10] &  0.7 & 0.9  \\ 
\hline
\end{tabular}
\tablefoot{
\tablefoottext{a}{Data from \cite{Vanderburg:2020aa}}
\tablefoottext{b}{Bad weather conditions}
\tablefoottext{c}{Target spectra close to bad column} 
}
\label{tbl:obs_log}
\end{table*}

The existence of planets can also be inferred from the characteristics of their debris disks or from their atmospheric pollution. Among the few dozen metal-polluted WDs studied so far, \citet{Manser:2019ab} detected a stable 123~min periodic variation in the Ca~{\sc ii} emission lines of the WD SDSSJ122859.93+104032.9; their interpretation involves a solid body with a size between 2 and 600~km at a distance of 0.73~R$_\odot$ under the assumption of a circular orbit. In a similar way,  the composition of the circumstellar gaseous disk of the hot WD J091405.30+191412.25 resembles the predictions for the deeper layers of icy giant planets \citep{Gansicke:2019aa}. In their scenario, a giant planet orbiting at about 15~R$_\odot$ would be undergoing evaporation of its atmosphere, which is then accreted onto the WD. 

On the target of this paper, \target, \citet[][V20 hereafter]{Vanderburg:2020aa} found transits in data from the TESS mission \citep{Ricker:2015aa}, indicating a planet candidate with a size comparable to Jupiter. With an orbital period of 1.4~d, its 8~min long grazing transits imply that a substantial fraction of the WD is occulted, reaching transit depths of $\approx$57\%. Nearly identical transit depths observed in the Sloan-$g$ filter and the Spitzer 4.5~$\mu m$ channel were used by V20 to provide an upper limit to the mass of the giant planet candidate of 14~M$_{\rm{Jup}}$. They determined the host WD as a nearby ($\sim$25~pc) cool ($T_\mathrm{eff}=4710 \pm 60$~K) stellar remnant, that is also a known member of a triple system, with two M-dwarfs separated by 58~AU orbiting the WD at a projected distance of $\approx$1500~AU. 

These companion M-dwarfs have been used to explain the evolution and current orbit of the planet candidate via the Zeipel-Lidov-Kozai mechanism (ZKL, \citealt{von-Zeipel:1910aa,Lidov:1962aa,Kozai:1962aa}); something that is explored in detail in three works: \citet{Munoz:2020aa} use the ZKL to constrain the primordial (during the main-sequence of the host star) semi-major axis and mass of the companion to 2-2.5~AU and 0.7-3~M$_{\rm{Jup}}$, respectively. Two other works obtain less constrained primordial parameters, with a distance of 10-20~AU \citep{OConnor:2021aa} or 20-100~AU and a preference for the larger values \citep{Stephan:2020aa} and in both cases without constraints on the orbiter's mass. \citet{Lagos:2021aa} argue that common envelope evolution is at least as plausible as the ZKL mechanism to explain the formation of \target\,b and they predict its mass to be larger than about 5~M$_{\rm{Jup}}$. Finally, V20 also present dynamical instabilities due to planetary multiplicity as another scenario that works more efficiently for low planet masses (super-Earths to Saturn masses, \citealt{Maldonado:2021aa}).

For cool WDs such as \target, it is not surprising to find spectra that show only a continuum, without distinguishable spectral lines, since the most common H and He transitions disappear below $\approx$5000~K and $\approx$12,000~K, respectively. This is the case of the spectra shown by V20 and the absence of lines also hampers the measurement of the mass of the companion through the Doppler effect.

However, an alternative method to constrain the mass of an exoplanet was proposed by \citet{de-Wit:2013aa}: By measuring the Rayleigh scattering slope in a transmission spectrum, it is possible to estimate the occulter's atmospheric scale height, and -- given the occulter's radius and some assumptions on the mean molecular mass of the atmosphere --, its mass. In the current work, we present high precision multi-band photometry and low resolution spectroscopy that was obtained with the 10.4m GTC telescope, with the aim to constrain the mass of the companion with that method. 

In the remainder of this paper, we describe in Sect.~\ref{sec:obs} the observations and data reduction of the different data sets. In Sect.~\ref{sec:wdparams} we derive the atmospheric parameters of the WD with two independent data sets: using archival photometric data, and using the H$\alpha$ absorption line that is detected for the first time in the GTC spectra. Sections~\ref{sec:bb_transits} and  \ref{sec:trans_spec} describe the analysis of the broad-band time series photometry and of the transmission spectroscopy. These are used in Sect.~\ref{sec:cons} to constrain the mass of the transiting object.  As a by-product, accurate transit timings of sub-second precisions were obtained; in Sect.~\ref{sec:omc} we refine the current ephemeris and look for deviations from linearity. Finally, in Sects.~4 and 5 we discuss the results and present our conclusions.

\section{Observations and data reduction}
\label{sec:obs}

\subsection{GTC photometry}

\begin{figure*}[ht]
\centering
\includegraphics[width=\textwidth]{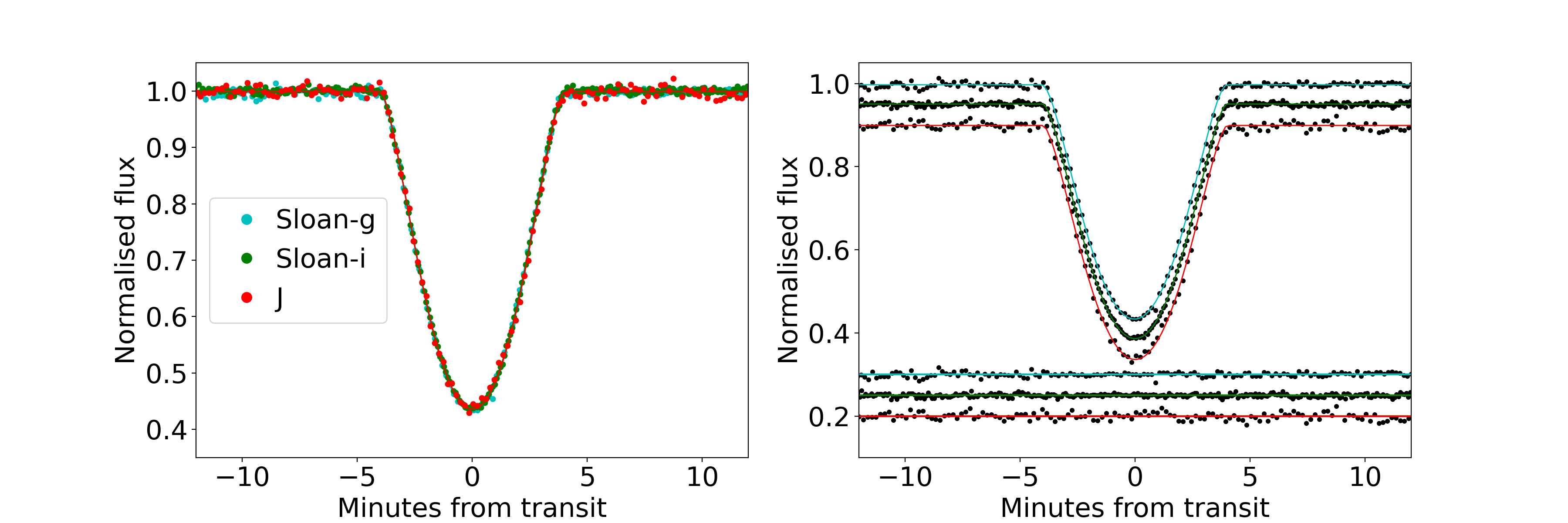}
\caption{Light curves of \target\ obtained with GTC in three different filters. In the right panel, the curves are plotted introducing vertical offsets to observe the cadence and precision of each of the data sets, and the residuals of the best fit model. The dispersion of the residuals in each color is 0.6, 0.4 and 0.9\%, respectively.} 
\label{fig:fig_ft}
\end{figure*}

We first observed \target\ during a transit event on the night of 2020 March 2, using the frame-transfer photometric mode of the GTC's OSIRIS instrument \citep{Cepa:2000aa} with a Sloan-$i$ filter (see Table~\ref{tbl:obs_log} for a list of observations and their configurations). The exposure time was 5~s, which is also the cadence of the light curve. We extracted the light curve with the HiPERCAM data reduction package \citep{Dhillon:2018aa}, using a variable aperture and one reference star to compute the differential and normalised light curve (Fig.~\ref{fig:fig_ft}). The dispersion of the residuals at the time of transit, after subtracting the transit model described in Sect.~\ref{sec:anal}, is 0.4\% per data point. For consistency, we also re-analised the GTC data from 2019 Oct. 21 presented in V20, which had been obtained with the same observing setup except for the Sloan-$g$ filter and exposure time. Using the same reduction procedure, the dispersion of the residuals in this case is 0.6\%. 

Furthermore, on 2020 June 13 we obtained $J$-band (centered at 1.25~$\mu m$) photometry using the GTC's EMIR instrument \citep{Garzon:2016aa}. The observing strategy was similar to \citet{Alonso:2008ab}: We obtained a series of images in the "stare" mode during a predicted transit, with an exposure time of 10~s in a pre-selected region of the detector with few cosmetic defects. To correct for the variable background emission -- which might arise from changes in the thermal background or in pixels sensitivities  -- we obtained sequences of images before and after the transit observation, in which the target is offset by $\approx$30\arcsec\ in a direction that avoids placing other bright objects in the region where the target was acquired during the transit. All the images were calibrated using standard procedures. The offset images before and after transit were averaged using sigma-clipping and the two resulting images were averaged to obtain a \emph{sky}-frame. This frame was subtracted from each of the images of the "stare" sequence. From them, we extracted the fluxes of the target and of the reference star using the same HiPERCAM data reduction package as in the visual-band images described above. After the fitting process described in Sect.~\ref{sec:anal}, the dispersion of the residuals is 0.9\%.

The extracted light curves in the three different bandpasses are shown in Fig.~\ref{fig:fig_ft}. The three light curves are remarkably similar, which is to be expected if the emission of the secondary component (i.e. the candidate planet) is negligible in all these bands.

\subsection{GTC spectroscopy}

Three more transits were observed on 2020 May 3, 10 and 20 using the spectroscopic mode of OSIRIS; the first one with the R1000R grism and the two later ones with the R1000B grism; see again Table~\ref{tbl:obs_log}. We used mostly a 2.5\arcsec slit, and rotated the slit to include both \target\ and a slightly brighter nearby star, identified as Gaia DR2 2146576727003500032, at a distance of $\approx$1\arcmin. 

The 2020 May 3 observations with the R1000R grism suffered from strongly variable and poor seeing conditions, and despite being analysed with the same procedures as described below, the results were not of sufficient quality, and we do not include them in the further analysis. During the 2020 May 10 observations, now with the R1000B grism, the spectrum of the target was accidentally placed very close to a bad column of the detector. We repeated the observations in the night of 2020 May 20, placing the spectrum at a more favourable location of the CCD. We used integration times of 75~s and a 500~kHz readout speed, resulting in a cadence of about 92~s. This provided a good signal during the transit spectra while also getting a reasonable sampling of the transit. 

After the regular calibration of the spectra taken on the night of 2020 May 10, we applied a spline interpolation at each row of the detector to correct for the bad column located at one of the wings of the target spectra. For the two nights, the spectra of both the target and the comparison star were calibrated, sky-subtracted and then extracted with {\sc starlink}/{\sc pamela}\footnote{\url{http://starlink.eao.hawaii.edu/starlink/} and \url{http://deneb.astro.warwick.ac.uk/phsaap/software/pamela/html/INDEX.html}} \citep{Marsh:1989aa}.

\begin{figure}
\centering
\includegraphics[width=0.5\textwidth]{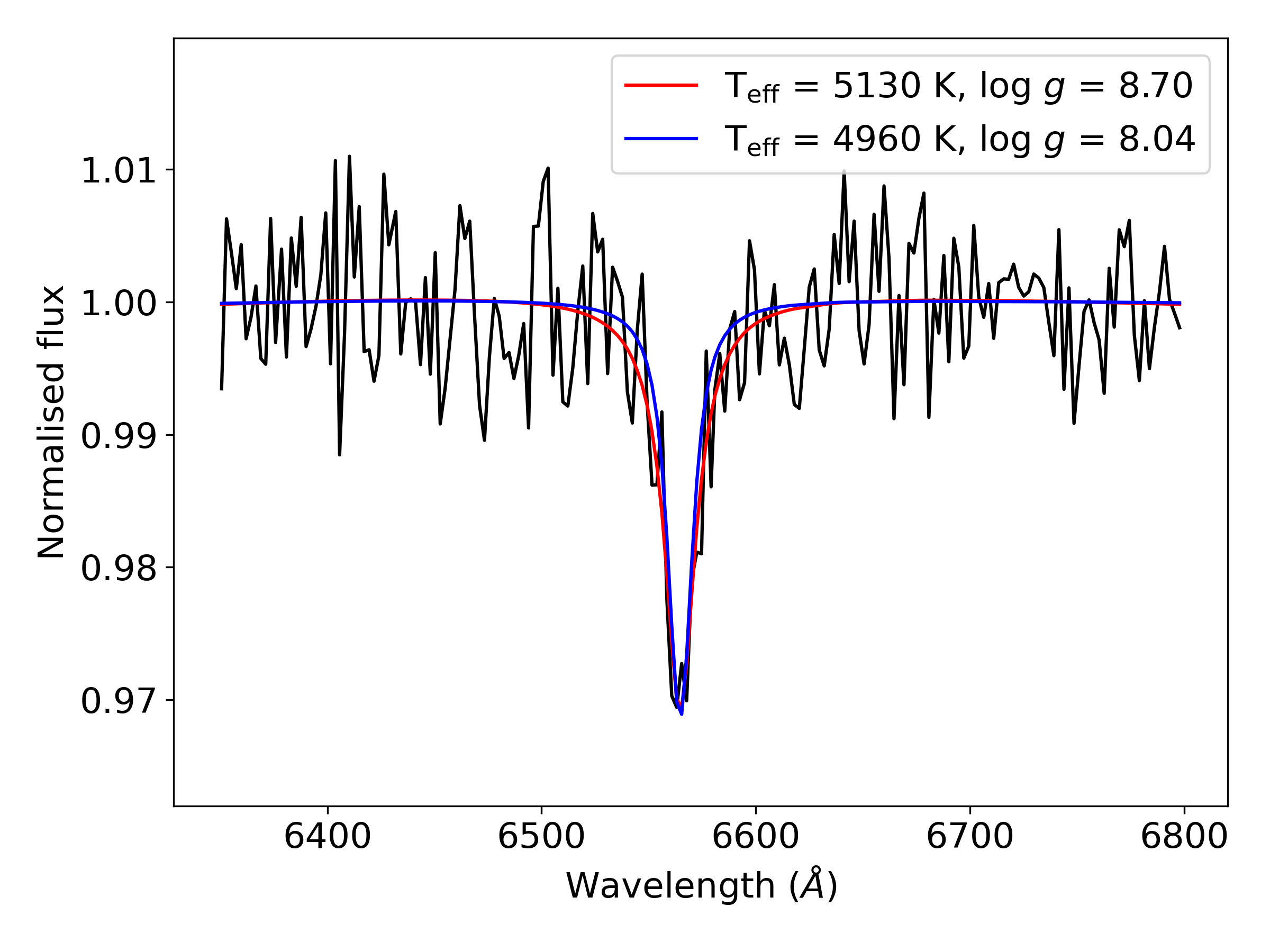}
\caption{The average out-of-transit normalised spectrum of the night of 2020 May 20, centred at the H$\alpha$ absorption line. The best-fit spectroscopic model of a DA WD is overplotted as a red line, and the model that best fits the photometric data as a blue line.} 
\label{fig:halpha}
\end{figure}

To obtain a high signal-to-noise off-transit reference spectrum of \target\, we computed an average spectrum from the 38 exposures taken out-of-transit on the night of 2020 May 20. In this average spectrum, the single and most prominent non-telluric feature is the H$\alpha$ absorption line, which is detected with a depth of about 3\% (Fig.~\ref{fig:halpha}). Given this detection, we consider \target\ to be of class DA, which is defined by the presence of Balmer lines and the simultaneous absence of other lines \citep{McCook:1999aa}.  While the H$\alpha$ absorption line is also detected in the out-of-transit spectra of the other nights, for the further analysis we used only the May 20 spectrum, so as to avoid potential systematics from the interpolation of the bad column or the bad weather conditions. 

\section{Data analysis}
\label{sec:anal}

\subsection{Parameters of the WD}
\label{sec:wdparams}

We re-determined the atmospheric parameters of \target\ using absolute photometry from archival sources. In addition, we obtained an independent estimation of the parameters from the modelling of the H$\alpha$ absorption line identified in the GTC spectra. In both cases, we used the Markov-Chain Monte Carlo (MCMC) {\sc emcee} package within {\sc python} \citep{Foreman-Mackey:2013aa} to fit a grid of model spectra to the two data sets. A synthetic grid of pure DA-WDs was computed using the code of \citet{Koester:2010aa}, spanning $T_\mathrm{eff}=4000-6000$\,K in steps of $250$\,K and $\log g=7.0-9.0$ in steps of 0.25\,dex. We explored the whole parameter space and minimised the \chisq\ with 30 chains and 150,000 walkers per chain, employing uniform priors for $\log g$ and $T_\mathrm{eff}$. Each pair of $T_\mathrm{eff}$ - $\log g$ corresponds to a mass $M_{\rm WD}$, radius $R_{\rm WD}$, and cooling age $\tau_{\rm{cool}}$ as derived from the evolutionary cooling sequences for DA WDs\footnote{\url{http://www.astro.umontreal.ca/~bergeron/CoolingModels}} (\citealt{Tremblay:2011aa,Blouin:2018aa,Bedard:2020aa}). 
 
\begin{table*}
\centering
\caption{Derived parameters of \target\ from the photometry (Pan-STARSS1, 2MASS, WISE 1 and 2 bands, \textit{Gaia} eDR3 blue and red bands), using different thresholds (Err$_{\rm{min}}$) for the minimum uncertainties in the input magnitudes.}
\begin{tabular}{lccc}
\hline
\hline
Parameter & Err$_{\rm{min}}$=0 & Err$_{\rm{min}}$=0.01 & Err$_{\rm{min}}$=0.05 (Adopted)\\
\hline
$T_\mathrm{eff} [K]$ & 4750$\pm$80 & 4705$\pm$90& 4960$\pm$100\\
$\log g$ & 7.87$\pm$0.04 & 7.86$\pm$0.05 & 8.04$\pm$0.06\\
$M_{\rm WD}$ [M$_\odot$] & 0.512$\pm$0.023& 0.506$\pm$0.034& 0.606$\pm$0.039\\
$R_{\rm WD}$ [M$_\odot$] & 0.0137$\pm$0.0004 & 0.0138$\pm$0.0004 & 0.0123$\pm$0.0005\\
$\tau_{\rm{cool}}$ [Gyr] & 5.9$\pm$0.7 & 6.0$\pm$0.8 & 6.4$\pm$1.0\\
$\chi^2_{red}$ & $\sim$50 & $\sim$9 & $\sim$2\\
\hline

\end{tabular}
\label{tab:phot_fits}
\end{table*}

The estimation of the atmospheric parameters of \target\ using absolute photometry requires the synthetic spectra to be scaled by the solid angle of the star, $\pi \times (R_{\rm WD}/D)^{2}$, with $D$ being the distance to the WD. The radius is computed for each pair of trial $T_\mathrm{eff}$ - $\log g$ as explained above. The distance was derived from the \textit{Gaia} early Data Release 3 (eDR3; \citealt{Gaia-Collaboration:2020aa,Gaia-Collaboration:2016aa}) parallax ($\varpi=40.39 \pm 0.05$\,mas) as $D=24.76\pm0.03$\,pc. To address the interstellar extinction affecting the photometric points we reddened the grid of models by adopting $E(B-V)=0.01$, which was determined from the 3D dust map produced by {\it Stilism}\footnote{\url{https://stilism.obspm.fr/}}\ \citep{Lallement:2018wc}. The scaled and reddened grid of models was then convolved with the bandpass of each photometric filter and compared with the observed fluxes in each filter. The latter had been converted from the published magnitudes using the zero-points available on the Spanish Virtual Observatory (SVO) Filter Profile Service\footnote{\url{http://svo2.cab.inta-csic.es/theory/fps/}}. Through the MCMC fit, we obtained the best $T_\mathrm{eff}$, $\log g$ combination that matches the observation and made use of the evolutionary cooling sequences to derive the $M_{\rm WD}$, $R_{\rm WD}$, and $\tau_{\rm{cool}}$ that correspond to those values. Several tests using different sets of photometric data points and distinct scaling values for the errors were  performed, during which we could also replicate the results obtained by V20, who used only the Pan-STARRS1 and 2MASS photometry. However, we decided to include all available photometry in the fit and adopted the solution from the use of the \textit{Gaia} eDR3, Pan-STARRS1, 2MASS, and WISE 1 and 2 bands. The best fit using the published magnitude uncertainties has a reduced \chisq\  of $\approx$ 50, suggesting a poor fit or underestimated errors. We then explored different solutions by setting minimum uncertainties in the photometric magnitudes of 0.01 and 0.05, which replaced the reported errors if they were smaller than these thresholds. These are summarised in Table~\ref{tab:phot_fits}. 

We obtained an independent estimation of the effective temperature and surface gravity from the H$\alpha$ absorption line. We continuum-normalised this line in both the observed and synthetic spectra, by fitting a low-order polynomial to the surrounding continuum. The synthetic spectra were degraded to the resolution of the observed data (1.98~\AA). The best-fit model was found for $T_\mathrm{eff}$=5130$\pm$55\,K and $\log$ g=8.7$\pm$0.2\,dex. However, we note that these are statistical uncertainties from the MCMC analysis, while systematics that are potentially larger may arise from the use of a single spectral line in the derivation of these parameters.

For the remainder of the paper, we adopt the values from the best-fit model to the photometric data, setting a minimum uncertainty of 0.05~mag, corresponding to the rightmost column in Table~\ref{tab:phot_fits}. Figure~\ref{fig:halpha} shows the best-fit models to the spectroscopic and photometric data. We can see that the model using our adopted values (blue line) is in a reasonable agreement with the observed H$\alpha$ absorption line, and is also within the uncertainties of the values reported in V20 using pure hydrogen models ($T_\mathrm{eff}=4785 \pm 60$\,K and $\log g=7.93 \pm 0.03$\,dex).

\subsection{Fit of the broad-band transits}
\label{sec:bb_transits}

We use the mass and radius of \target\ derived above as normal priors to the fit of the three transits observed in broad-band photometry. To perform this fit and to obtain the posterior parameter distributions, we used the {\sc exoplanet} package \citep{dan_foreman_mackey_2020_3785072}. The other parameters of the fit were the transit epoch $T_{\rm c}$, the mean baseline flux $f_{\rm out}$, the two coefficients $u_{\rm b}$ and $u_{\rm c}$ of a quadratic limb darkening law, with the sampling proposed by \citet{Kipping:2013aa}, the radius of the secondary $R_{\rm p}$, and the impact parameter $b$. The finite integration times were taken into account with a factor of 5 oversampling on the fitted models. After fitting the maximum a posteriori parameters, we sampled the posteriors using the NUTS (No U-Turns Step method, \citealt{hoffman2014no}) implementation of the {\sc python pyMC3} package\footnote{\url{https://docs.pymc.io/}}. We ran two chains with 4,000 steps for tuning and 5,000 draws for a final chain length of 10,000 for each of the parameters. We assessed the convergence of the chains with the \citet{Gelman:1992aa} test, that was below 1.002 for each parameter. All three transits were analysed independently using the same procedure. In Table~\ref{tab:params_3colors} we provide the mean and standard deviation of the derived parameters, and in Fig.~\ref{fig:fig_ft} we show the best fit models. Note that the input parameter $b$ is much larger than 1, since this parameter is defined relative to the (small) radius of the WD. In the Appendix we provide the corner plots of the posteriors to check for the correlations between the different parameters.

\begin{table*}
\centering
\caption{Derived parameters of \target\ b in the three high-cadence light curves, and assuming a circular orbit.}
\begin{tabular}{lccc}
\hline
\hline
Parameter & Sloan $g$ & Sloan $i$ & $J$ \\
\hline
$R_{\rm p}$ [R$_\odot$] &  0.104$\pm$0.006 & 0.110$\pm$0.007 & 0.111$\pm$0.007\\
$b$ & 8.3$\pm$0.8 & 8.9$\pm$0.9 & 9.0$\pm$0.9\\
$u_{\rm b}$  & 0.13$\pm$0.10 & 0.23$\pm$0.16 & 0.25$\pm$0.20\\
$u_{\rm c}$ & 0.08$\pm$0.15 & 0.37$\pm$0.30 & 0.24$\pm$0.29\\
$f_{\rm out}$ & 0.9994$\pm$0.0004& 1.0007$\pm$0.0003 & 0.9992$\pm$0.0007 \\
$R_{\rm p,fix}$ [R$_\odot$]\tablefootmark{a} & 0.105975$\pm$0.000014 & 0.105981$\pm$0.000016 &  0.105974$\pm$ 0.000039 \\
\hline

\end{tabular}
\tablefoot{
\tablefoottext{a}{Radius of the secondary assuming $b$=8.5, $R_{\rm{WD}}$=0.0123~R$_\odot$, and with the quadratic limb darkening coefficients $u_{\rm b}$ and $u_{\rm c}$ fixed to the values given by \citet{Gianninas:2013aa} for a DA WD with $\log$ g=8.0~dex and $T_{\rm eff}$ = 5000~K. For the values of these coefficients in the $J$ filter, we extrapolated the \citet{Gianninas:2013aa} values using polynomial fits, obtaining $u_{\rm b}$=-0.027 and $u_{\rm c}$=0.353. } 
}
\label{tab:params_3colors}
\end{table*}

\subsection{Transmission spectroscopy}
\label{sec:trans_spec}

For the study of a potential dependence of the radius of the secondary against wavelength, we make some assumptions: Regardless of the 'real' size ($R_{\rm WD}$) and mass ($M_{\rm WD}$) of the host, we assume that these values are identical at every wavelength, and we fix them to 0.0123~R$_\odot$ and 0.606~M$_\odot$, respectively (Sect.~\ref{sec:wdparams}). A second assumption is that the impact parameter $b$ is also independent of the wavelength, and we fix it to 8.5. Selecting different values for these quantities (within the ranges given in Tables~\ref{tab:phot_fits} and~\ref{tab:params_3colors}) has an effect on the slope of the transmission spectrum that lies within the slope's uncertainty. Finally, the two quadratic limb darkening coefficients at each wavelength bin were fixed to results of a spline fit to the coefficients of a DA WD with $T_{\rm eff}$ = 5000~K and $\log g$ = 8.0~dex, provided by \citet{Gianninas:2013aa}.

\begin{figure}
\centering
\includegraphics[width=0.5\textwidth]{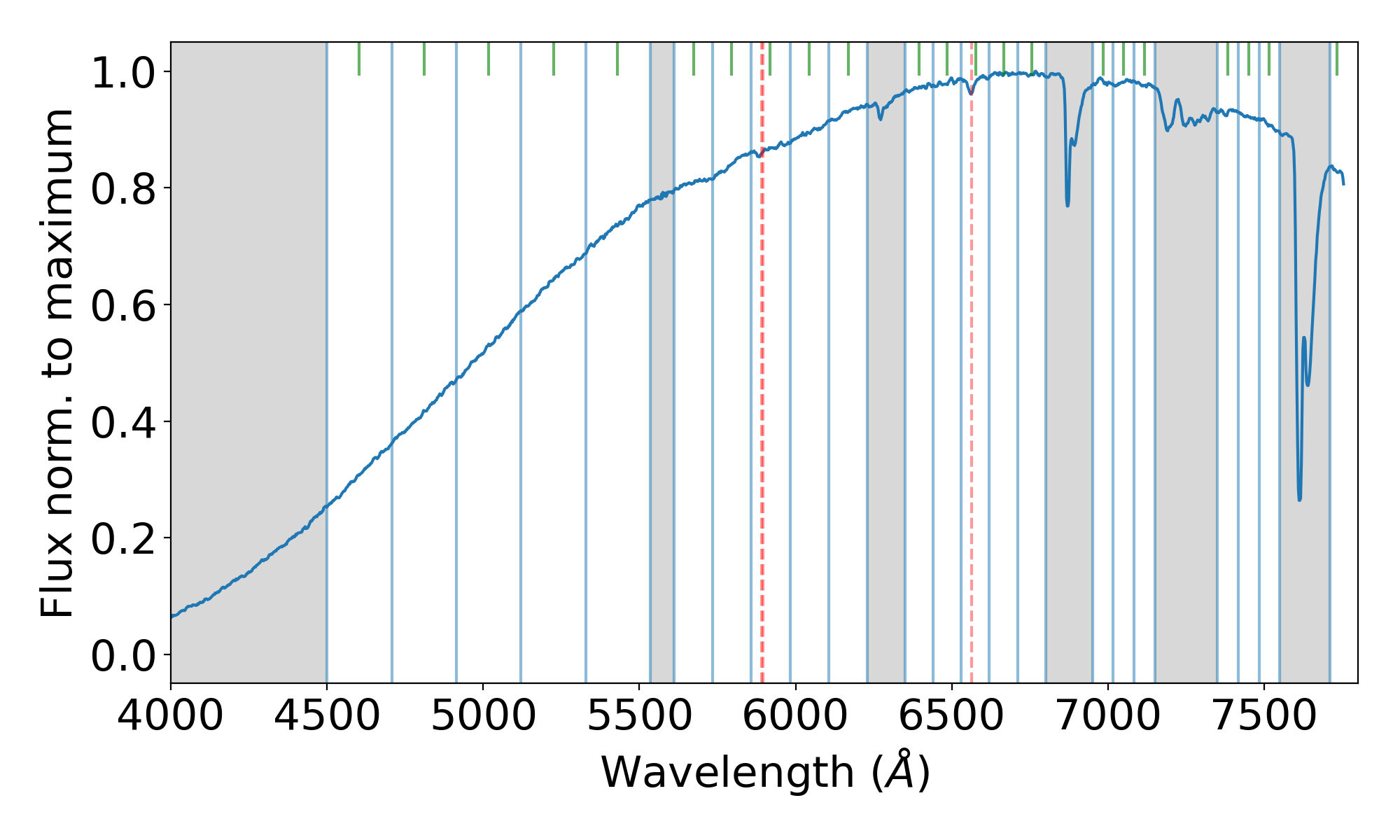}
\caption{The average out-of-transit spectrum of \target\ with the wavelength bins used to compute the transmission spectrum. The limits of the bins are represented with vertical blue lines, whereas the central wavelengths of each bin are indicated on the top with short vertical green lines. The definition of the bins took into account the locations of the telluric absorption bands (marked with the gray bands; these are discarded), and narrower bins were used when the flux was higher. The positions of the Na~{\sc i} doublet and the H$\alpha$ absorption line are marked with vertical dashed red lines. } 
\label{fig:color_bins}
\end{figure}

We divided the spectra of the target in bins of different sizes as shown in Fig.~\ref{fig:color_bins}. These bins were selected to avoid the wavelengths affected by telluric absorptions and to be narrower when the captured signal was higher. Some fine-tuning of the number of bins between two strong telluric absorptions was performed with the aim of completely including the H$\alpha$ and Na~{\sc i}  lines inside single bins. We constructed a light curve for each wavelength bin by summing the flux of the target, the flux of the comparison star, dividing both, and normalising to the average out-of-transit flux.

At each wavelength bin, we fit for the radius of the secondary $R_{\rm p}$, the mean out-of-transit flux, and a slope to account for potential systematic trends in the data. The rest of the parameters are held fixed as described above. The sampling of the posterior distributions was done using the same formalism as in the previous section. To include the broad-band transit photometry, we also performed a fit were we kept as free parameters  only $R_{\rm p}$, $T_{\rm c}$ and the mean and slope of the out-of-transit flux, while fixing $b$, the mass and radius of the WD, and the limb darkening coefficients.

\begin{figure*}
\centering
\includegraphics[width=\textwidth]{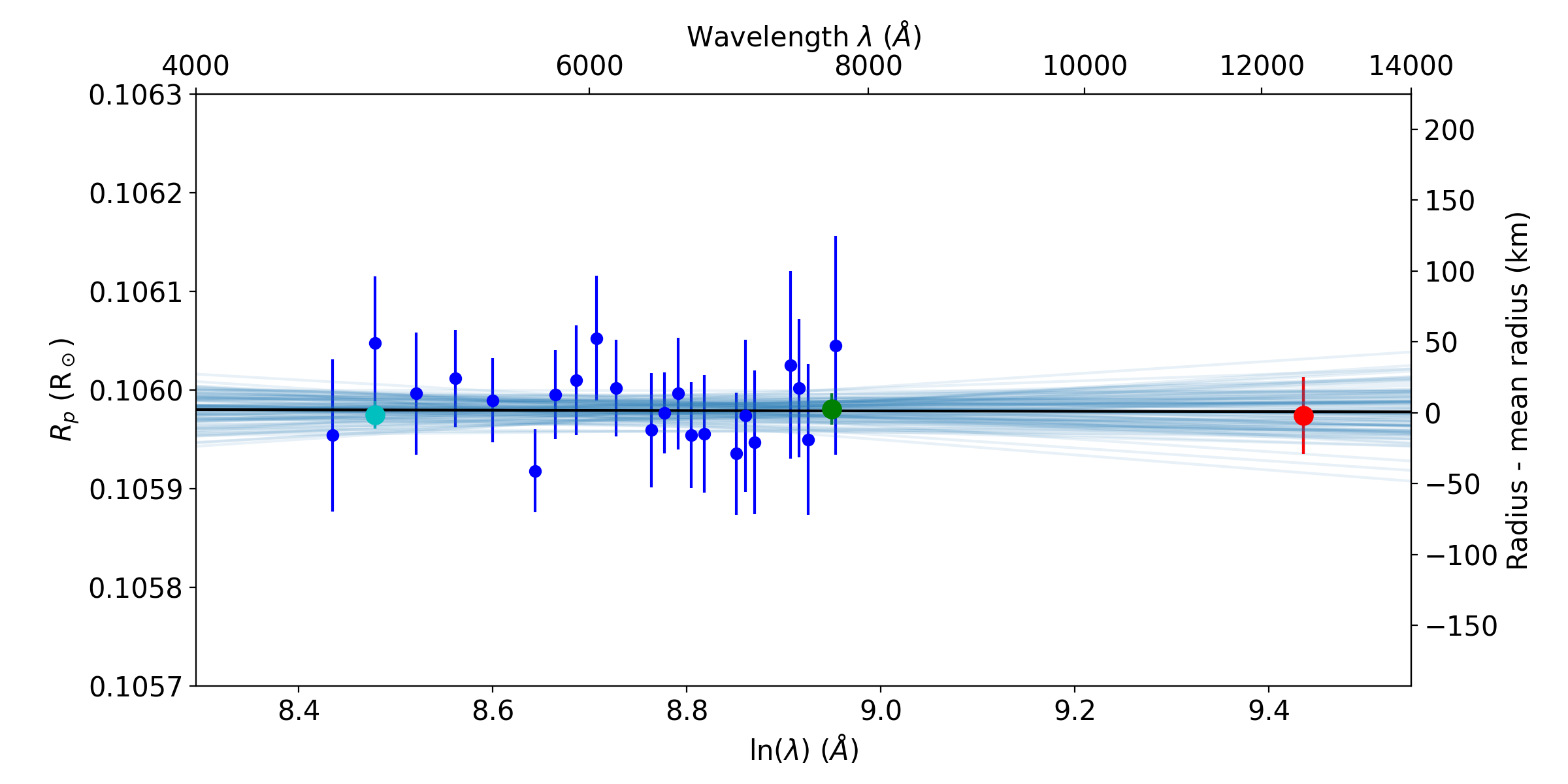}
\caption{The transmission spectrum of \target\ b. The radius of the WD is fixed to the estimated average value (0.0123~R$_\odot$), and the impact parameter is held fixed at a given value (8.5), as both parameters are independent of the wavelength. The larger filled circles mark the radius of the secondary in the Sloan $g$ (cyan), Sloan $i$ (green), and $J$ (red) filters, while  dark blue dots mark the radii determined at each spectral bin defined in Fig.~\ref{fig:color_bins}. The solid blue lines in the background are random samples of the posterior of the linear fits to measure the atmospheric scale height. Note the logarithmic scale in the x-axis. The vertical scale on the right side gives the difference in km from the mean radius of the secondary.} 
\label{fig:trans_spec1}
\end{figure*}

The 2020 May 20 transmission spectrum (Fig.~\ref{fig:trans_spec1}) does not show evidence of any spectral features that are dependent on the atmospheric scale height. Moreover, the spectrum is essentially flat, which is in agreement with the thee data points corresponding to the transits observed in broad-band filters. This is an indication of either a low atmospheric scale height of the secondary, -- which implies that it is a relatively massive body --, or of a  masking of its spectral features due to having a thick cloud deck at high altitudes. In the next section, we use the broadest feature expected in giant planets, the Rayleigh scattering, to place a lower limit to the mass of the secondary. We repeated the analysis with the transit observed in 2020 May 10, in which we attempted to correct for the effects from the bad detector column in the spectra. The transmission spectrum for that night, shown in the Appendix (Fig.~\ref{fig:trans_spec2}), is noisier at every wavelength bin, and appears to show systematically larger planetary radii that are marginally inconsistent with the photometric data points. This is indicative of residual systematic effects, and we decided not to include this spectrum in the further analysis.

\subsection{Constraints on the mass of the secondary}
\label{sec:cons}

\begin{figure}
\centering
\includegraphics[width=0.5\textwidth]{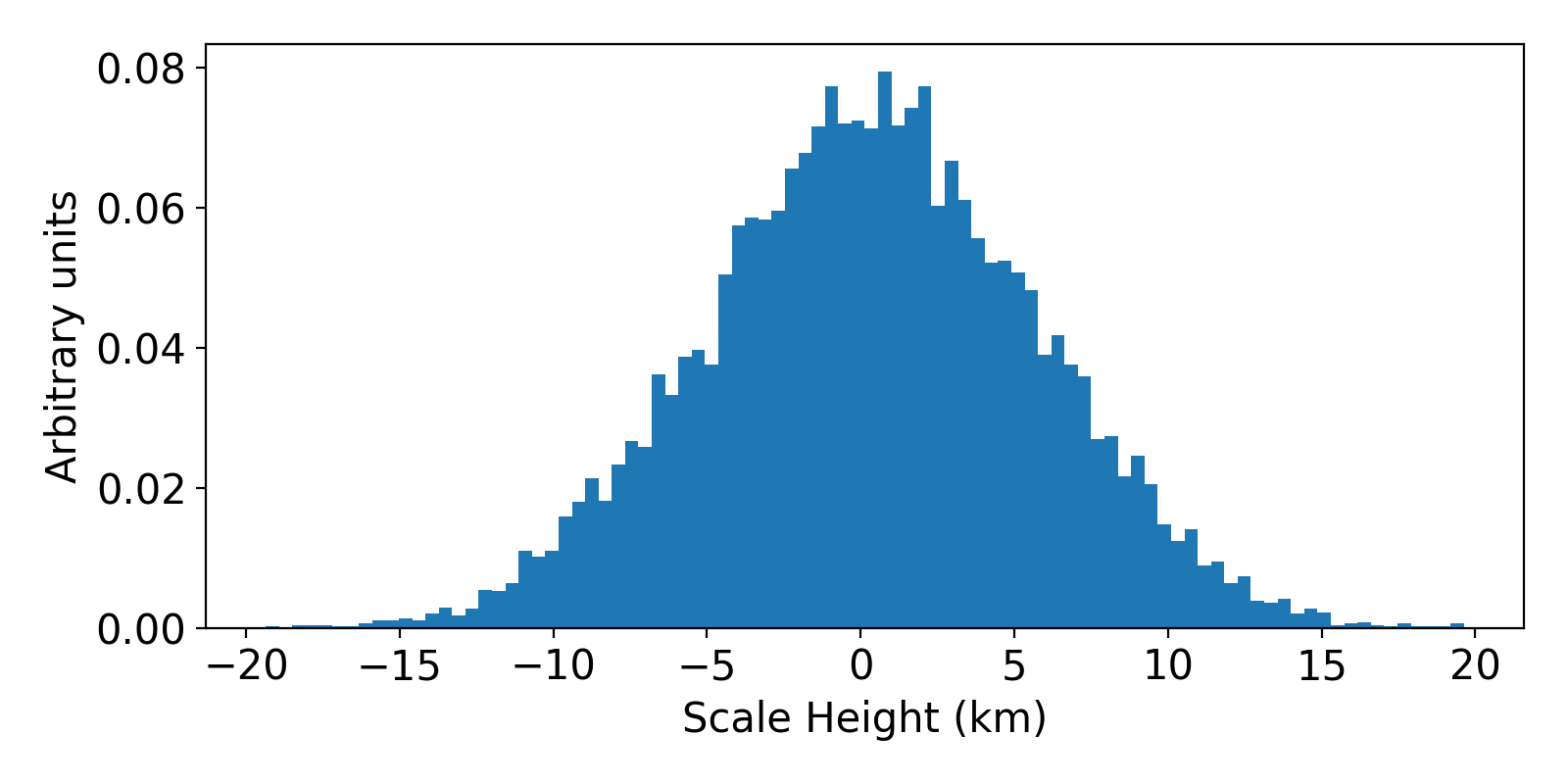}
\caption{The histogram of the scale heights obtained from the fits to the slope of the transmission spectrum shown in Fig.~\ref{fig:trans_spec1}.} 
\label{fig:scale_height}
\end{figure}

The mass of a gas giant planet can be estimated from the slope of its transmission spectrum assuming that it is due to Rayleigh scattering in the occulter's atmosphere: the slope is proportional to the atmospheric scale height $H=\frac{kT}{\mu g}$, with $k$ the Boltzmann's constant, $T$ the temperature of the atmosphere, $g$ the surface gravity and $\mu$ the mean molecular mass ($\approx$2.3 for a H/He dominated atmosphere). Following  \citet{de-Wit:2013aa}, the scale height for a given slope $\frac{d R_{\rm p}(\lambda)}{d ln\lambda}$ is then derived from:
\begin{equation}
\alpha H = \frac{d R_{\rm p}(\lambda)}{d ln\lambda}~,
\end{equation}
where the factor $\alpha$ describes the dependence of the cross section of the main absorbent against wavelength; in the case of Rayleigh scattering this is $\alpha=-4$. The mass of the occulting body $M_{\rm p}$ is then given through:
\begin{equation}
M_{\rm p}=\frac{k T R_{\rm p}^2}{\mu G H}~,
\label{eq:mp}
\end{equation}
with $G$ being the gravitational constant. We performed a linear fit to the transmission spectrum in Fig.~\ref{fig:trans_spec1} to obtain the slope, and thus the atmospheric scale height assuming a Rayleigh scattering regime. We included data from both the broad-band imaging and the low resolution spectra taken on 2020 May 20. For $T$, we took the equilibrium temperature assuming a Bond albedo of 0.35 (as in V20), a circular orbit, and isotropic re-emission of the incident flux, resulting in 165~K, in agreement with the values obtained by V20. The posterior distribution of the obtained scale heights is shown in Fig.~\ref{fig:scale_height}. It shows a maximum compatible with a null slope and discards scale heights larger than 5.8, 11, and 16~km at the 1-, 2-, and 3-$\sigma$ levels, respectively. With Eq.~\ref{eq:mp} and the $R_{\rm p,fix}$ from Table~\ref{tab:params_3colors}, these heights can be translated to corresponding minimum masses of 4.4, 2.4, and 1.6~$M_{\rm Jup}$.

We revised the values that would arise from relaxing the assumptions that enter into these fits. First, we explored the posterior distribution of the mass when the secondary's estimated temperature is at the lowest reasonable level, by assuming a Bond albedo of 0.6 and a $R_{\rm WD}$ at the lower end (0.0118~R$_\odot$), resulting in $T=143$~K. The smaller value of $R_{\rm WD}$ in this grazing configuration implies a larger inferred $R_{\sc p}$ of 1.11~R$_{\rm{Jup}}$ with a higher impact parameter to match the total duration of the transit (see the corner plots in the Appendix). With these assumptions, the 1-, 2-, and 3-$\sigma$ levels of the maximum scale heights and minimum masses are 6.3, 11, and 17~km, or 4.0, 2.3, and 1.5~M$_{\rm Jup}$, respectively. An exploration of the higher end value of $R_{\rm WD}$ (0.0128~R$_\odot$) implies a smaller $R_{\sc p}$ of 0.95~R$_{\rm{Jup}}$, which under the same assumptions as before results in $T = 149$~K, and 1-, 2-, and 3-$\sigma$ levels of the maximum scale heights and minimum masses of 4.1, 9.6, and 15~km, or 4.8, 2.0, and 1.3~M$_{\rm Jup}$. From these tests we can conclude that the effect of relaxing the assumptions has only minor consequences on the determined lower mass limits for the secondary companion.

\subsection{Ephemeris improvement}
\label{sec:omc}

\begin{figure}
\centering
\includegraphics[width=0.5\textwidth]{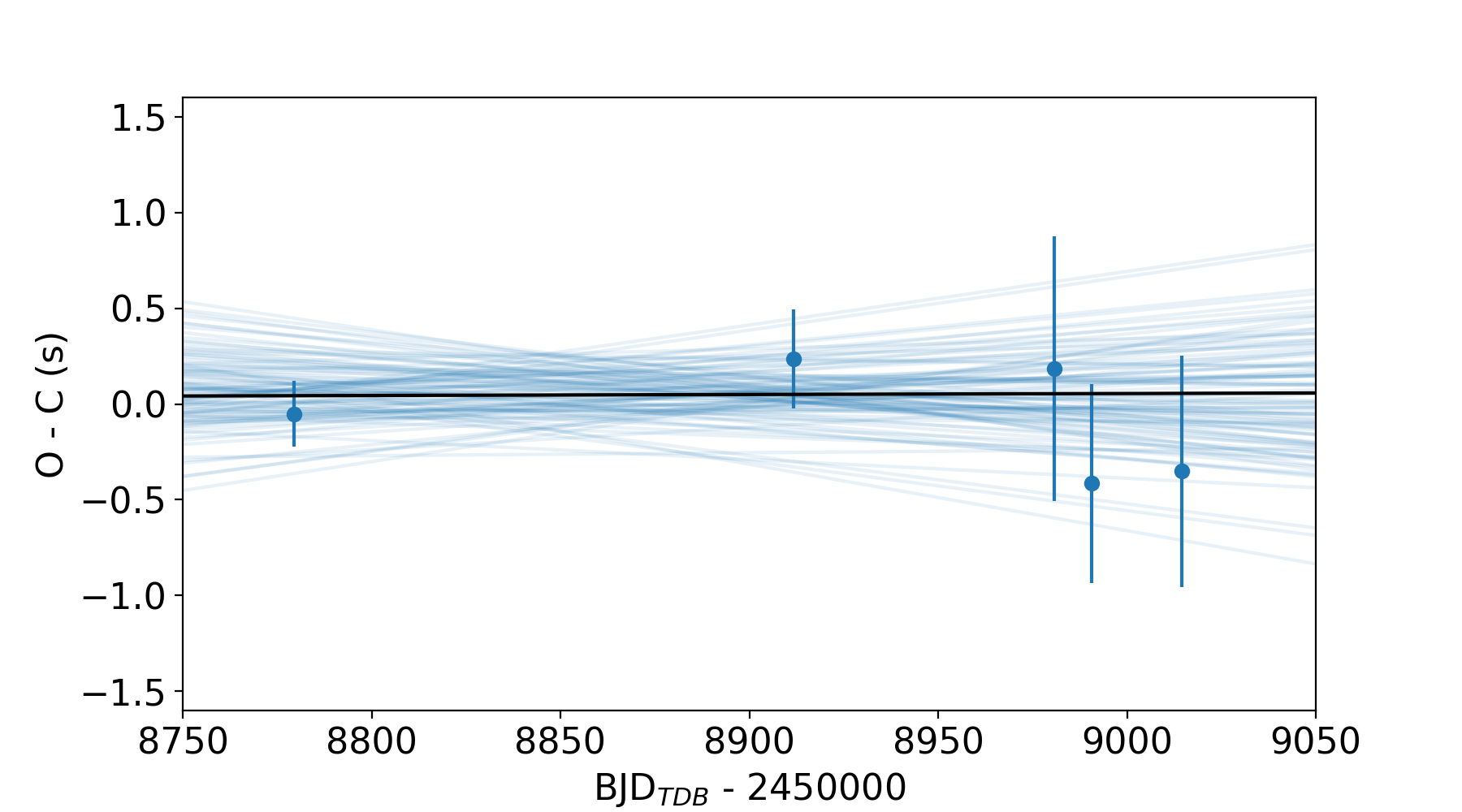}
\caption{The Observed minus Calculated (O-C) values of the mid-transit times for which a precision of less than 1~s could be achieved. The lines are random samples of the posterior of a linear fit to these values.} 
\label{fig:omc}
\end{figure}

Precise measurement of the mid-transit epochs are a widely used method to search for dynamical interactions between different planets in planetary systems (see \citealt{Agol:2018aa} for a review). An additional effect is the R\o mer or Light-Time Effect (LITE) due to the movement of the barycenter of a system of more than two bodies (e.g. \citealt{Irwin:1959aa}). With the transits of \target\ b being deep and short, it is possible to accurately measure the central times with a precision that is comparable to the one achieved for post-common envelope binaries (e.g. \citealt{Marsh:2018aa}). As part of the fitting process in the previous sections, the times of mid-transit have been measured with an accuracy of less than 1\,s for all GTC observations, except for the one affected by bad weather. The epochs and corresponding errors are given in Table~\ref{tab:epochs}. We use these times to refine the ephemeris of V20 to:
\begin{equation} 
T_0 (BJD_{TDB}) = 2458779.375086[2] + E \times 1.40793913[2]~,
\end{equation}
where the numbers in brackets indicate the uncertainty of the last digit. The corresponding Observed minus Calculated (O-C) diagram is shown in Fig.~\ref{fig:omc}.

\begin{table}
\centering
\caption{Measured mid-transit epochs of \target}
\begin{tabular}{llcl}
\hline
\hline
Epoch  & 1$\sigma$ & 1$\sigma$ & Source \\
(BJD$_{TDB}$ - 2450000)&(d)&(s)&\\
\hline
8779.375085 & 0.000002 & 0.2&GTC Sloan $g$ \\
8834.284788 & 0.000047 &4.1 &Spitzer 4.5$\mu$m \\
8911.721367 & 0.000003 & 0.3&GTC Sloan $i$ \\
8973.670701 & 0.000063 & 5.4 &GTC R1000B \\
8980.710383 & 0.000008 & 0.7&GTC R1000R \\
8990.565959 & 0.000006 & 0.5&GTC R1000R \\
9014.500916 & 0.000007 & 0.6&GTC EMIR $J$ \\
\hline

\end{tabular}
\label{tab:epochs}
\end{table}

\section{Discussion}
\label{sec:disc}

The observed transmission spectrum of \target\,b is consistent with being flat. This is an indication of either a cloud layer at low pressure levels respectively at high altitudes that masks any absorption or scattering feature, or of a low atmospheric scale height, indicating a secondary companion more massive than a few Jupiter masses.

The equilibrium temperature of \target\,b of $T \approx 165\pm20$~K is the second lowest known for transiting Jupiter-sized exoplanets, about 30~K warmer than the long-period planet Kepler-167e \citep[$P = 1071$~day,][]{Kipping:2016ab}, and around 50~K warmer than Jupiter. 

Efforts to obtain Jupiter's transmission spectrum by observing its shadow projected on one of its moons (\citealt{Montanes-Rodriguez:2015aa,Lopez-Puertas:2018aa}) showed that the optical part of the transmission spectrum of Jupiter is dominated by extinction from aerosols. Rayleigh scattering from the gas in Jupiter was a minor contributor to that transmission spectrum, as shown in the Fig. 3 of \citet{Lopez-Puertas:2018aa}. In the polar regions of Jupiter, both in-situ probes and ground-based observations have revealed a high altitude haze layer that is best understood with fractal aggregates of monomers with an effective radius of about 0.7$\mu m$ (\citealt{West:1991vk,Zhang:2013wj}). These hazes are thought to be similar in Titan, for which detailed NIR (0.88 to 5~$\mu$m) and far-UV (0.11 to 0.19~$\mu$m) transmission spectra were obtained during solar occultations of NASA's Cassini spacecraft (\citealt{Robinson:2014vs} and \citealt{Tribbett:2020uv}, respectively). 
 \citet{Robinson:2014vs} analysed the NIR spectra using a simple haze extinction model with a slope of -1.9$\pm$0.2. This slope is smaller than one induced by pure Rayleigh scattering (-4). \citeauthor{Robinson:2014vs} explain this slope from complexities of the scattering by haze particles, when they are in an intermediate regime between the limits of Rayleigh scattering and geometrical optics.  If the same type of hazes is present in our transmission spectrum and assuming a slope of -2, this would translate into scale heights larger by a factor of about 2, and the secondary's lower mass limits would be $\approx$2 times smaller.

Cloud-decks are also frequently invoked to explain flat transmission spectra; e.g. for super-earth planets, \citet{Howe:2012vn} show that cloud decks at the 10~mbar level or higher effectively suppress the characteristic slope of the Rayleigh scattering. Cloud-decks in Jupiter are however confined to pressures of 0.5-1.2-bars or larger, whereas its haze layer exist at pressures of $\approx50$~mbar at low latitudes and $\approx20$~mbar at high latitudes \citep{Guerlet:2020ub}. For cool exoplanets, hazes are therefore more likely to affect transmission spectra in the optical range, and to 'drown out' the slopes from Rayleigh scattering from the gas.
 
\begin{figure}
\centering
\includegraphics[width=0.5\textwidth]{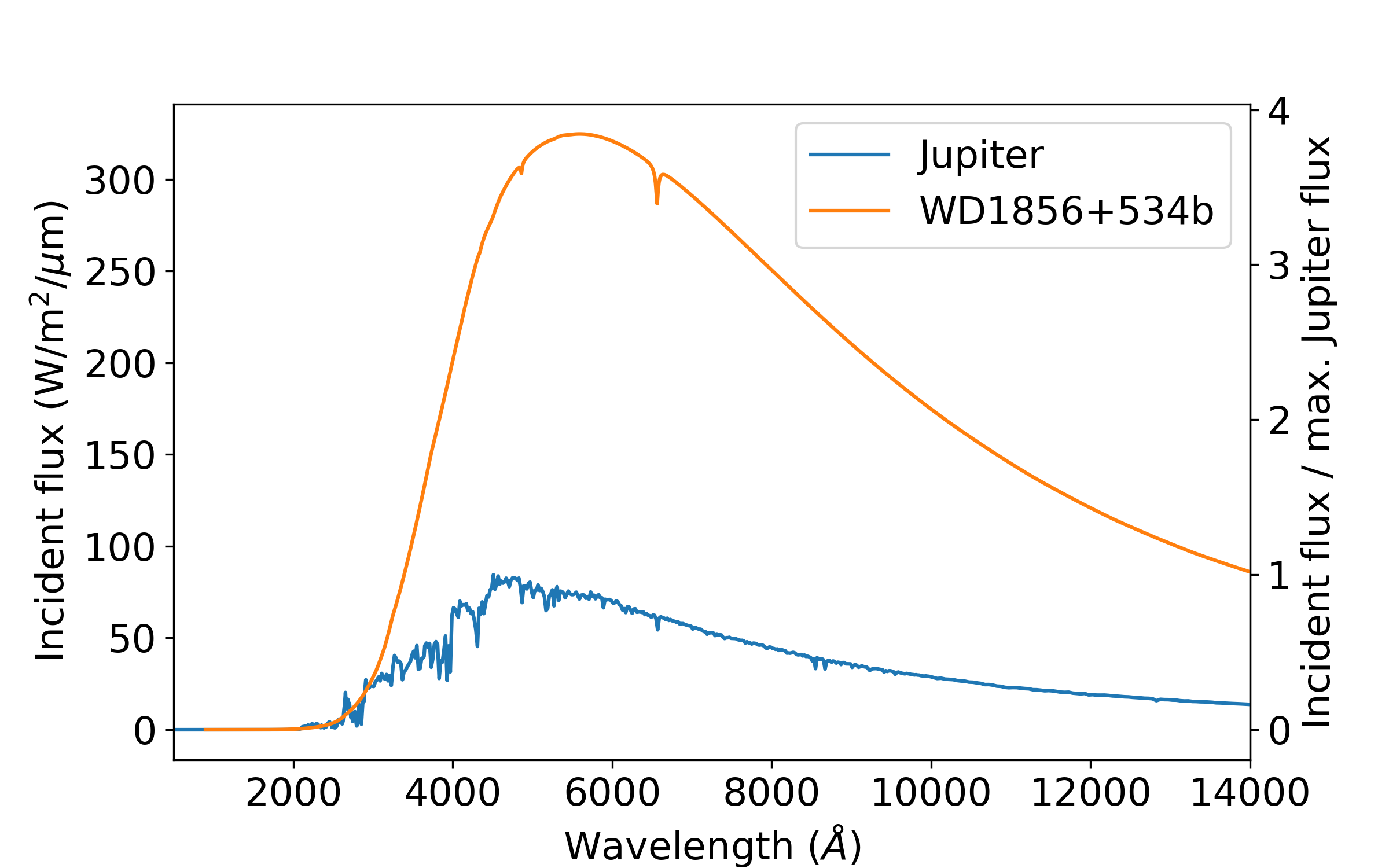}
\caption{The incident fluxes at the orbits of Jupiter and \target\,b.} 
\label{fig:jup_wd1856compar}
\end{figure}

The high altitude hazes at the polar region of Jupiter are thought to be created through photochemical effects induced by the precipitation of energetic particles in its auroral region \citep{Pryor:1991ua}, which could be more important than the effect of the incident solar UV flux. In the case of \target\,b, the incident UV flux is at the level of few times the solar flux received at Jupiter (Fig.~\ref{fig:jup_wd1856compar}), and the grazing nature of its transits imply that the region sampled is most likely a polar one. An estimation of the probability, location, and intensity of auroras in \target\,b is out of the scope of this paper, but it seems reasonable, from the incident UV flux alone, to assume that high altitude hazes might be present in its atmosphere.

\citet{Karalidi:2013vj} numerically studied the spectropolarimetric signals expected in reflected light from gaseous exoplanets, and found that polar hazes will not be detectable in exoplanets studied in integrated reflected light: the difference between a model with polar hazes and one without is negligible in the disk-integrated flux. Thus, the transits of \target\,b might provide an opportunity to reveal these hazes with more precise future observations. For that purpose, a detailed modelling with all the physical processes at play (extinction, atmospheric refraction, gas absorption, multiple scattering by cloud and haze particles,...) would have to prove that a haze signature can be disentangled from the other atmospheric effects \citep{Robinson:2014vs}. In the Solar System, inaccurate considerations on the role of these processes have led to historically conflicting or wrong estimations of gaseous abundances and locations of clouds (see \citealt{West:2018wf} for a recent review).

Assuming a clear atmosphere, with a mean molecular mass similar to Jupiter, and dominated by scattering processes (either Rayleigh scattering on the H$_2$-He gas or from a potential haze), our results are indicative of a companion to the WD with a few Jupiter masses. Among the proposed formation and evolution scenarios for \target\,b (see Sect.~\ref{sec:intro}), the common envelope evolution under the conditions explored in \citet{Lagos:2021aa} is more efficient for secondaries of several Jupiter masses -- they predict the mass to be above 5 M$_{\rm Jup}$ -- while the mechanisms favouring lower-mass secondaries, such as the ZKL effect or the dynamical instabilities due to planet-planet scattering are more difficult to accommodate.  

We have shown that the mid-transit epochs of \target\,b can be measured with precisions of a fraction of a second. Using the LITE equations from \citet{Irwin:1959aa}, a further 5 [10] M$_{\rm Jup}$ companion orbiting with a period of 1000 [500] days would translate into O-C amplitudes of 7 [8] seconds, with the amplitude increasing with the third body's orbital period. More massive companions would have been detected through the spectral energy distribution (SED) or through the dilution of the transits in the Spitzer data, as has already been explored in V20. The detection of an eventual light-time effect would require long-term observations of transits with timing precision comparable to those presented here, similar to observing campaigns that have been performed for compact evolved eclipsing binary stars over several decades \citep{Marsh:2018aa}. Given the short duration of the transits, this would however not be very demanding in terms of required telescope time.

\section{Conclusions}

We have obtained multi-color and low-resolution spectroscopic observations of six transits of \target\,b, obtaining a transmission spectrum that is essentially flat.  Under the assumption that the transmission spectrum is dominated by an undetected Rayleigh scattering, the absence of a significant slope allows us to obtain lower limits to its mass of 4.4, 2.4, and 1.6~M$_{\rm Jup}$ at the 1-, 2-, and 3-$\sigma$ levels, respectively. These mass limits would favour a common envelope origin for the current orbit of \target\,b and disfavour other proposed mechanisms such as ZKL or dynamical scattering from additional planets. As an alternative, high-altitude haze layers may be present, which would imply flatter-sloped transmission spectra, and corresponding mass limits would be few times smaller. High clouds could also cause a flat transmission spectrum, but in a cold object like \target\,b, such a spectrum is more likely dominated by hazes.

The average out-of-eclipse spectrum shows a previously undetected H$\alpha$ absorption in the atmosphere of the host WD, that indicates a DA type. We fit a grid of synthetic spectra of DA WDs to this single line and obtain values of its effective temperature and gravity that are in reasonable agreement with those obtained from an analysis of the SED. We have additionally refined the ephemeris of the transits, and have shown that the achieved precision allows for a search of other potential giant planets in the system by means of the light time effect during a longer term observing campaign.

\begin{acknowledgements}
We thank the anonymous referee for useful comments that helped to improve the paper.  We thank F. Murgas for kindly providing the GTC Sloan $g$ raw data used in V20. RA acknowledges funding from the Spanish Research Agency (AEI) of the Ministry of Science and Innovation (MICINN) and the European Regional Development Fund (FEDER) under the grant PGC2018-098153-B-C31. PR-G acknowledges funding from the same agencies under grant AYA2017-83383-P and HJD under grants ESP2017-87676-C5-4-R and PID2019-107061GB-C66, DOI: 10.13039/50110001103. PI acknowledges financial support from the Spanish Ministry of Economy and Competitiveness (MINECO) under the 2015 Severo Ochoa Programme MINECO SEV–2015–0548. This publication made use of VOSA, developed under the Spanish Virtual Observatory project supported from the Ministry of Science and Innovation (MICINN) through grant AYA2017-84089. HiPERCAM is funded by the European Research Council under the European Union's Seventh Framework Programme (FP/2007-2013) under ERC-2013-ADG Grant Agreement no. 340040 (HiPERCAM). This research made use of \textsf{exoplanet} \citep{dan_foreman_mackey_2020_3785072} and its
dependencies \citep{Agol:2020ab, Kumar:2019aa, Astropy-Collaboration:2013aa,Astropy-Collaboration:2018aa,
Kipping:2013aa, Luger:2019aa, exoplanet:pymc3, The-Theano-Development-Team:2016aa}. This work has made use of data from the European Space Agency (ESA) mission {\it Gaia} (\url{https://www.cosmos.esa.int/gaia}), processed by the {\it Gaia} Data Processing and Analysis Consortium (DPAC,\url{https://www.cosmos.esa.int/web/gaia/dpac/consortium}). Funding for the DPAC
has been provided by national institutions, in particular the institutions participating in the {\it Gaia} Multilateral Agreement. We acknowledge the use of Tom Marsh's {\sc pamela}.

\end{acknowledgements}

\bibliographystyle{aa}
\bibliography{references_v0}

\begin{appendix}

\section{corner plots}

\begin{figure*}
\centering
\includegraphics[width=0.9\textwidth]{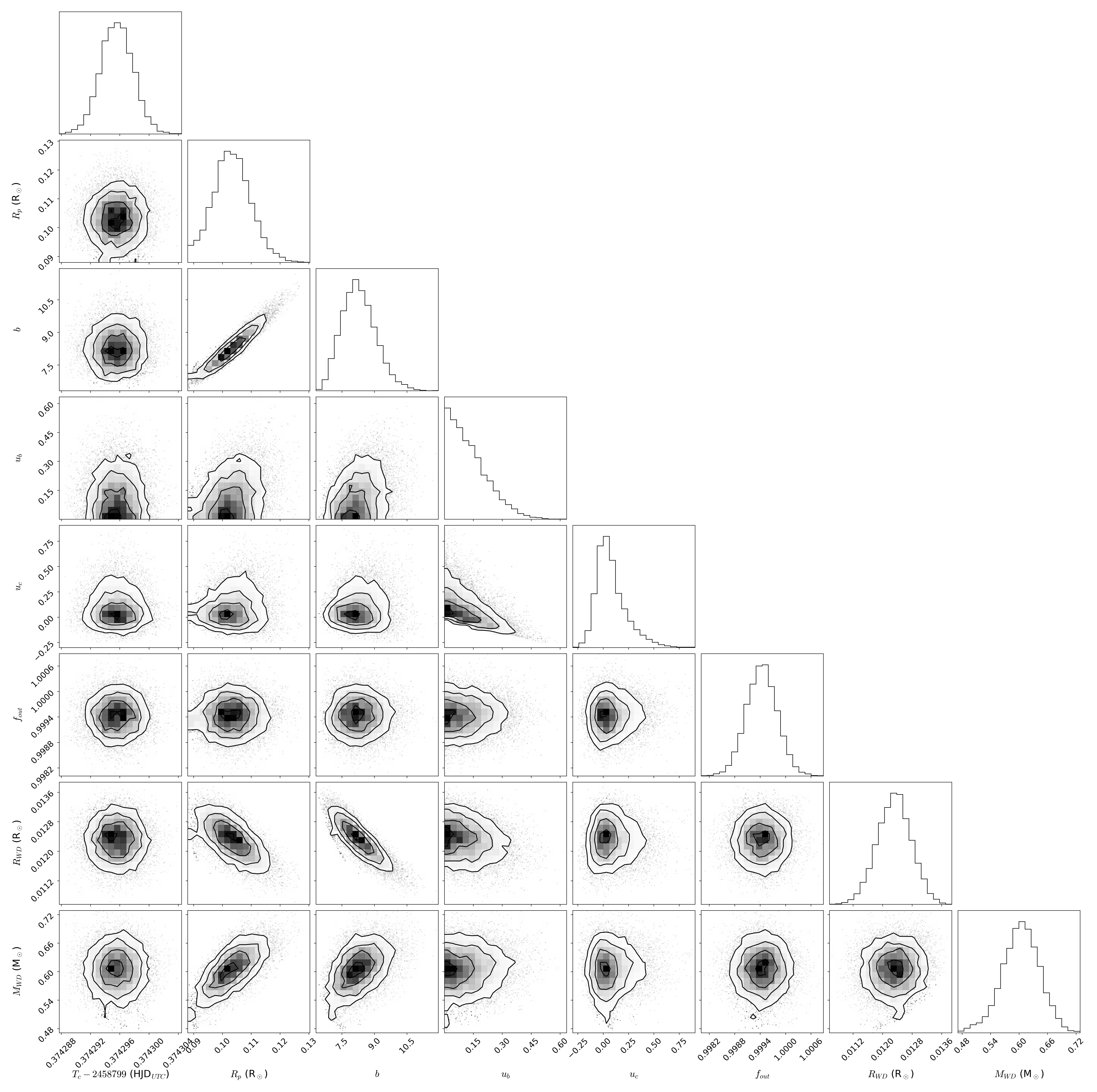}
\caption{Corner plots of the fit to the broad-band Sloan-$g$ photometry.} 
\label{fig:corner}
\end{figure*}

\begin{figure*}
\centering
\includegraphics[width=\textwidth]{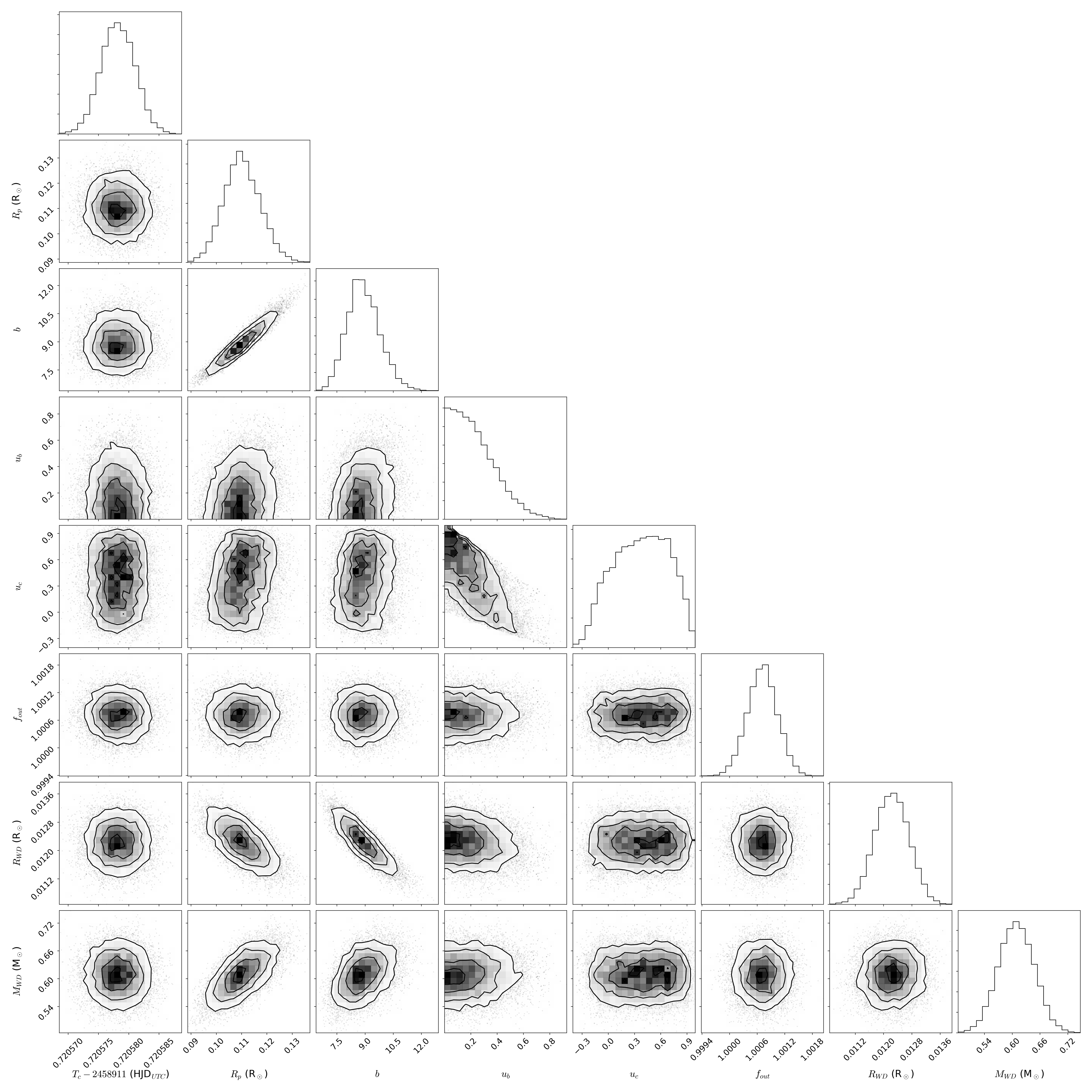}
\caption{Corner plots of the fit to the broad-band Sloan-$i$ photometry.} 
\label{fig:corner2}
\end{figure*}

\begin{figure*}
\centering
\includegraphics[width=\textwidth]{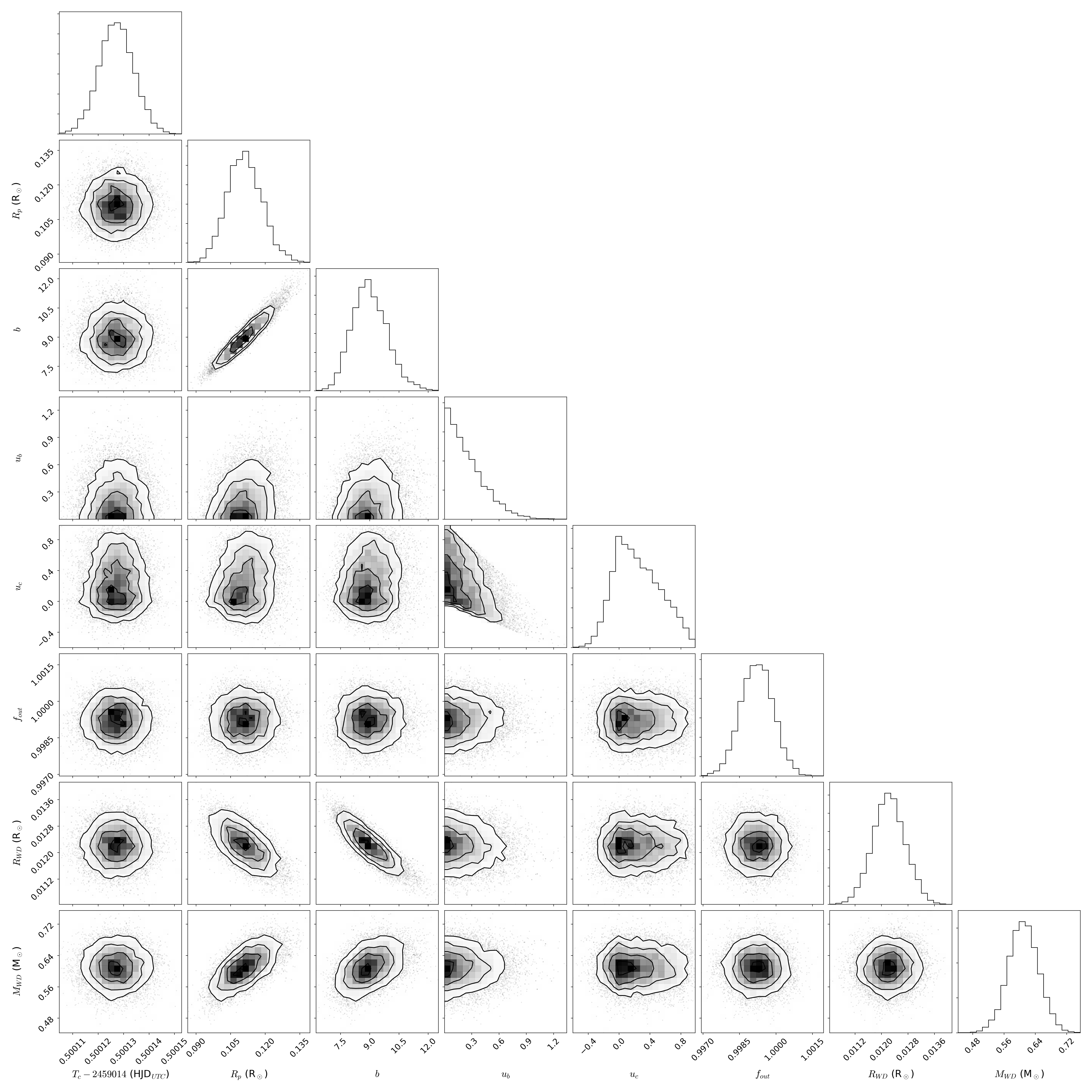}
\caption{Corner plots of the fit to the broad-band EMIR-$J$ photometry.} 
\label{fig:corner3}
\end{figure*}

\begin{figure*}
\centering
\includegraphics[width=\textwidth]{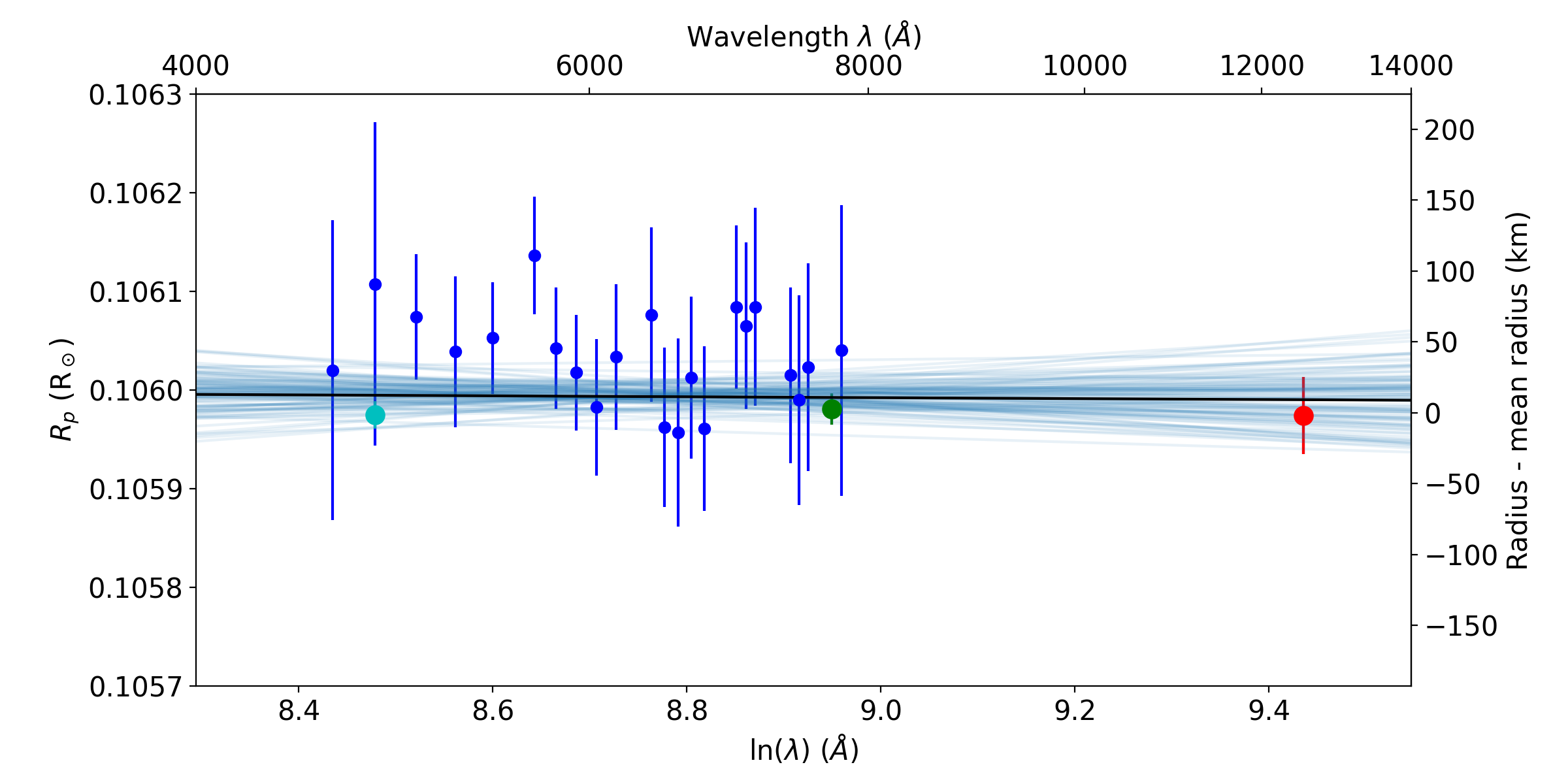}
\caption{Same as Fig.~\ref{fig:trans_spec1} for the night of 2020 May 10.} 
\label{fig:trans_spec2}
\end{figure*}

\end{appendix}

\end{document}